\documentclass[fleqn,10pt]{wlscirep}
\usepackage[utf8]{inputenc}
\usepackage[T1]{fontenc}
\usepackage{graphicx}
\usepackage{dcolumn}
\usepackage{bm}
\usepackage{epsfig}
\usepackage{subfig}
\usepackage{placeins}
\usepackage{graphicx}
\usepackage{amsmath}
\usepackage{amssymb}
\usepackage{svg}
\graphicspath{{Fig/}}
\usepackage[labelfont=bf,justification=RaggedRight,format=plain]{caption}
\usepackage[ruled,vlined]{algorithm2e}
\usepackage{algpseudocode} 
\usepackage{threeparttable}
\usepackage{multirow}

\title{Deep learning of material transport in complex neurite networks}

\author[1]{Angran Li}
\author[1,2,3,4]{Amir Barati Farimani}
\author[1,2,*]{Yongjie Jessica Zhang}
\affil[1]{Department of Mechanical Engineering, Carnegie Mellon University, Pittsburgh, PA, USA}
\affil[2]{Department of Biomedical Engineering, Carnegie Mellon University, Pittsburgh, PA, USA}
\affil[3]{Department of Chemical Engineering, Carnegie Mellon University, Pittsburgh, PA, USA}
\affil[4]{Department of Machine Learning, Carnegie Mellon University, Pittsburgh, PA, USA}

\affil[*]{jessicaz@andrew.cmu.edu}

\begin{abstract}
    Neurons exhibit complex geometry in their branched networks of neurites which is essential to the function of individual neuron but also brings challenges to transport a wide variety of essential materials throughout their neurite networks for their survival and function. While numerical methods like isogeometric analysis (IGA) have been used for modeling the material transport process via solving partial differential equations (PDEs), they require long computation time and huge computation resources to ensure accurate geometry representation and solution, thus limit their biomedical application. Here we present a graph neural network (GNN)-based deep learning model to learn the IGA-based material transport simulation and provide fast material concentration prediction within neurite networks of any topology. Given input boundary conditions and geometry configurations, the well-trained model can predict the dynamical concentration change during the transport process with an average error less than 10\% and $120 \sim 330$ times faster compared to IGA simulations. The effectiveness of the proposed model is demonstrated within several complex neurite networks.
\end{abstract}
\begin{document}

\flushbottom
\maketitle
%
%
\thispagestyle{empty}


\section*{Introduction}\label{section:introduction}

The geometry of neurites is known to exhibit complex morphology which is essential for neuronal function and biochemical signal transmission. However, the highly branched networks of neurites also make it challenging to mediate the intracellular material transport because the material synthesis and degradation in neurons are carried out mainly in the cell body \cite{segev2000untangling,swanger2013dendritic} which leads to a long-distance transport for the material. The disruption of this long-distance transport can induce neurological and neurodegenerative diseases like Huntington’s, Parkinson’s and Alzheimer’s disease \cite{de2008role,gunawardena2005polyglutamine, millecamps2013axonal, kononenko2017retrograde, zhang2018modulation}. Therefore, it has attracted considerable amount of attention in recent years to study the transport mechanisms and build mathematical transport models in neurite networks. Recent studies show that the molecular motors play fundamental roles in the intracellular transport to carry the material and move directionally along the cytoskeletal structure like microtubules or actin filaments \cite{vale2003molecular,Hirokawa2010,franker2013microtubule}.

Motivated by these findings, different mathematical models based on partial differential equations (PDEs) have been proposed to help understand the transport mechanisms and the pathology of neuron diseases. For instance, Smith and Simmons developed a generic model of the molecular motor-assisted transport of cell organelles and vesicles along filaments \cite{smith2001models}. Based on this motor-assisted transport model, Friedman and Craciun presented an axonal transport model by considering the ability of material to bind more than one motor protein \cite{friedman2005model}. Craciun \textit{et al.} introduced the pausing of the material during the transport process in the model to study the slow axonal transport \cite{craciun2005dynamical}. In addition, several PDE models were developed to study the transport impairment by accounting for the traffic jam \cite{Kuznetsov2009} and microtubule swirls \cite{kuznetsov2010modeling} in abnormal neurons.
Though the aforementioned models provide reasonable mechanistic explanation of the transport mechanism or the formation of the transport impairment in neurites, most of these models were solved only in one-dimensional (1D) domain without considering the impact of complex neurite geometry. Recent advances in numerical methods and computing resources allow us to simulate detailed cellular process in complex geometry using 1D or three-dimensional (3D) PDE models. For instance, computational softwares based on finite element method (FEM) have been used to solve PDE models in neuron \cite{hines1997neuron} and cell biological systems \cite{loew2001virtual}. However, these tools have accuracy and computational cost issues when tackling highly branched geometry like neurite networks. Based on the conventional FEM, isogeometric analysis (IGA) \cite{HUGHES20054135} was proposed to directly integrate geometric modeling with numerical simulation and achieve better accuracy and robustness compared to FEM. With the advances in IGA, one can simulate the transport process by solving the PDEs in an accurately reconstructed neuron geometry. In our previous study, we developed an IGA-based simulation platform to solve a 3D motor-assisted transport model in several complex neurite networks and obtain the spatiotemporal material distribution \cite{li2019isogeometric}. However, the high computational cost to perform 3D simulations has limited its application in the biomedical field when fast feedback from the computer simulation is needed.
\begin{figure*}[!htb]
    \centering
    \includegraphics[width = \linewidth]{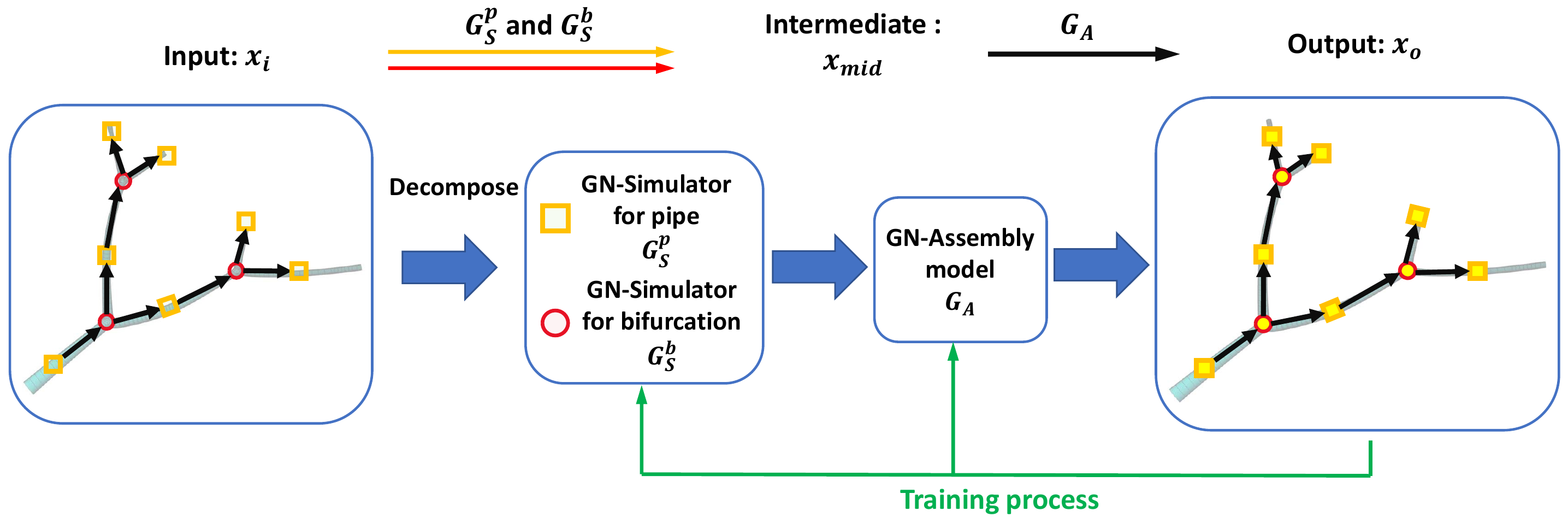}
    \vspace{-3mm}
    \caption{An overview of the GNN model to learn and predict neuron material transport process. The input neurite network is first decomposed into pipes and bifurcations to create the graph representation of the neurite network. Next, the input features $x_i$ of each pipe or bifurcation are processed by the corresponding GNN simulator ($G^p_S$ or $G^b_S$) to generate the intermediate concentration result $x_{mid}$. Then, the GNN assembly model ($G_A$) takes $x_{mid}$ as input and compute the interaction between simulators to predict concentration result $x_o$. }
    \label{fig:GNNFramework}
    \vspace{-5mm}
\end{figure*}

To address the limitation in the current simulation platform, we propose to build a surrogate model by combining deep learning (DL) with IGA simulation. DL has been proven successful in computer vision and image recognition \cite{krizhevsky2012imagenet, lecun2015deep, witten2016data} by handling large amount of labeled datasets and providing fast end-to-end prediction. The practical success of DL in artificial intelligence also inspires its application in solving high-dimensional PDEs \cite{han2018solving} and learning the physics behind PDE models \cite{raissi2019physics}. In particular, deep neural networks (DNNs) are becoming popular in surrogate modeling because it can be far more efficient when predicting complex phenomena \cite{he2019learning}. For instance, Farimani \textit{et al.} applied the conditional generative adversarial network (\textit{cGAN}) in a data-driven paradigm for rapid inference, modeling and simulation of transport phenomena \cite{farimani2017deep}. Wiewel \textit{et al.} proposed a DL-based fluid simulator to predict the changes of pressure fields over time \cite{wiewel2019latent}. Li \textit{et al.} developed an encoder-decoder based convolutional neural network (CNN) to directly predict the concentraion distribution of a reaction-diffusion system, bypassing the expensive FEM calcuation process \cite{li2020reaction}. While these works manage to learn the underlying physical models for prediction, they are limited to handle the problem in relatively simple geometry with Euclidean data (e.g. structured grid) available for training. Recently, many studies on graph neural networks (GNNs) have emerged to extend DL techniques for data defined in graph structure. Inspired by the success of CNNs \cite{krizhevsky2012imagenet}, a large number of methods were developed to re-define the convolutional operation for graph data and achieve great performance in computer vision tasks like graph node classification and image graph classification \cite{kipf2016semi,hamilton2017inductive,monti2017geometric,fey2018splinecnn}. GNNs were also applied to
predict the drag force associated with laminar flow around airfoils \cite{ogoke2020graph} or the property of crystalline materials \cite{karamad2020orbital},
understand the interaction behind physics scenes \cite{battaglia2018relational}, solve PDEs by modeling spatial transformation functions \cite{alet2019graph}, or learn particle-based simulation in complex physical systems \cite{sanchez2020learning}.

In this study, we develop a GNN-based model to learn the material transport mechanism from simulation data and provide fast prediction of the transport process within complex geometry of neurite networks. The use of GNN is motivated by the extensive topologies of neurite networks and the IGA simulation data stored in mesh structure. To ensure the model is applicable to any neurite geometry, we build the graph representation of the neuron by decomposing the neuron geometry into two basic structures: pipe and bifurcation. Different GNN simulators are designed for these two basic structures to predict the spatiotemporal concentration distribution given input simulation parameters and boundary conditions. Specifically, we add the residual terms from PDEs to instruct the model to learn the physics behind the simulation data. To reconstruct the neurite network, a GNN-based assembly model is used to combine all the pipes and bifurcations following the graph representation. The loss function of the assembly model is designed to impose consistent concentration results on the interface between pipe and bifurcation. The well trained GNN model can provide fast and accurate prediction of the material concentration distribution, leading to an efficient DL framework for studying the transport process in complex 3D models of neurite network. The framework was applied to several complex neurite networks achieving an average error less than 10\% and $120 \sim 330$ times faster compared to IGA simulations.

\section*{Results}
\subsection*{GNN model overview}
The aim of our GNN model is to learn from the material transport simulation data and predict the transport process in any neurite network. However, the geometry diversity of the neurite networks makes it impossible to train the DL model directly on the entire neurite network. To address this issue, we introduce the graph representation of the neurite network and build the GNN model based on the graph network (GN) framework \cite{battaglia2018relational}. A neurite network can be decomposed into two basic structures: pipe and bifurcation (yellow square and red circle in Fig. \ref{fig:GNNFramework}, respectively). Each structure can be treated as one node in the graph and the nodes can be connected following the skeleton of the neurite network to constitute the graph. Based on the graph representation of the neurite network, two separate GNN simulators are built for the pipe ($G_S^p$) and the bifurcation ($G_S^b$), respectively. Given input features $x_i$ including node locations, simulation parameters and initial nodal concentration, the simulators can output the intermediate nodal concentration result $x_{mid}$ in the pipe and the bifurcation, respectively. To obtain a consistent global concentration result, a GNN assembly model ($G_A$) is used to learn the interaction between different structures so that given intermediate value $x_{mid}$, the model can assemble different structures and output the final prediction $x_o$ on the graph. In sum, our GNN model consists of
\begin{itemize}
    \item The GNN simulators for local prediction in pipe and bifurcation structures; and
    \item The GNN assembly model for global prediction in the entire neurite network.
\end{itemize}
We implement our GNN model using PyTorch \cite{NEURIPS2019_9015} and the “PyTorch Geometric” library \cite{fey2019fast}. The detailed results will be explained in the following sections.

\subsection*{GNN simulators for pipe and bifurcation}
\begin{figure*}[!htb]
    \centering
    \includegraphics[width = \linewidth]{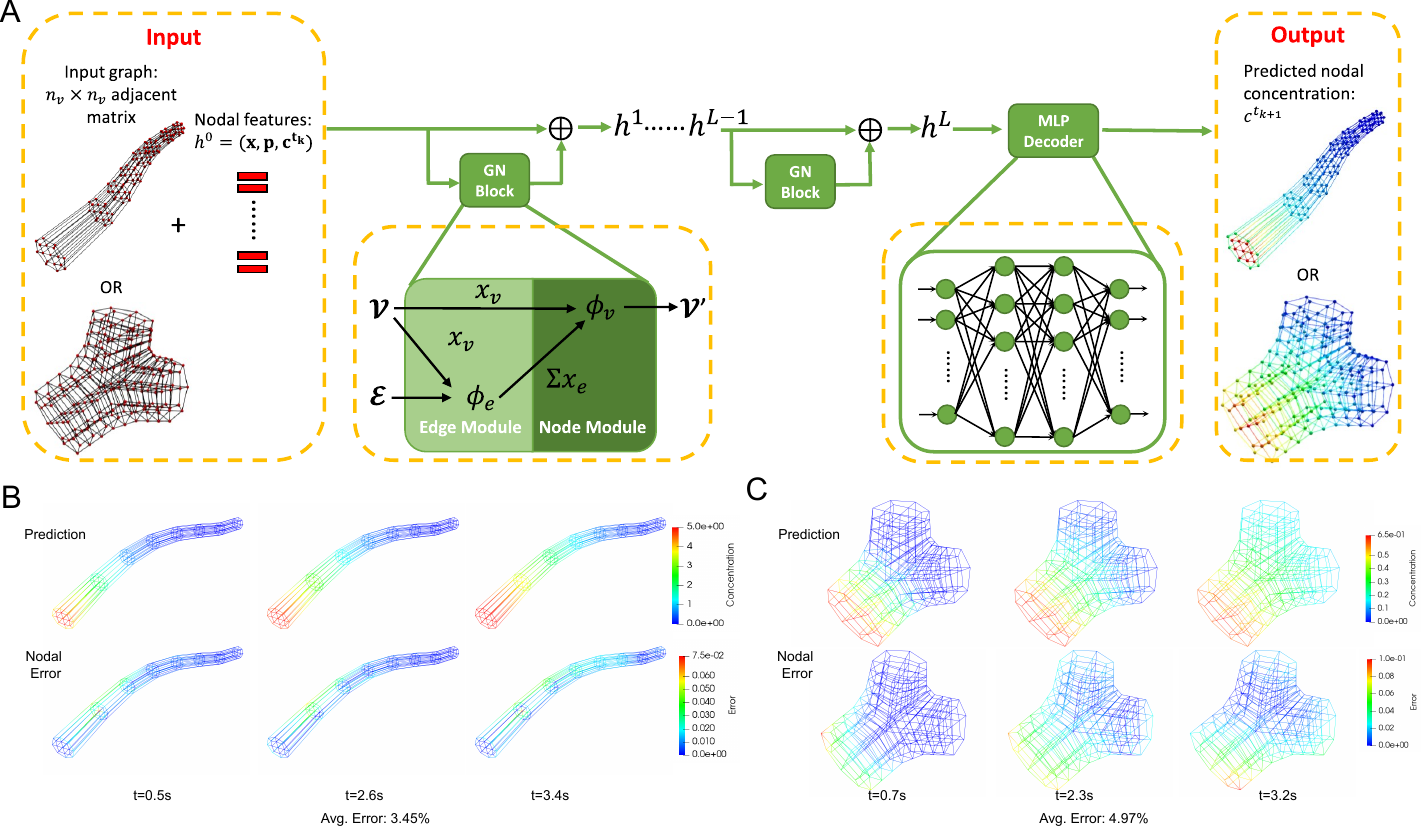}
    \caption{Results of GNN simulators. (A) The architecture of the GNN simulators adopts a recurrent ``GN Block + MLP Decoder" scheme on the input graph to compute nodal concentration values at each time step from input nodal features. The input $n_v$-node graph is stored in a $n_v \times n_v$ adjacent matrix. The nodal features include coordinate vector $x$, simulation parameter vector $p$ and initial concentration values $c^{t_k}$ at time step $t_k$. The $L$-step ``GN Block" follows Algorithm \ref{alg:GNblock} to compute the interaction among nodes and generates a series of updated latent graphs with hidden nodal embeddings, $h^1$, ..., $h^L$. The ``MLP Decoder" outputs nodal concentration values $c^{t_{k+1}}$ at the next time step from the nodal embedding $h^L$ of the final latent graph. (B, C) The concentration distribution comparison of pipe and bifurcation simulators, respectively. For each simulator, we plot the prediction and nodal error results at three different time steps. The average errors are 3.45\% and 4.97\% for the pipe and bifurcation simulators, respectively.}
    \label{fig:GNNSimulator}
    \vspace{-3mm}
\end{figure*}

Since the pipe and the bifurcation have different geometry topologies, we train two separate simulators to handle different graph structures extracted from the simulation results. Both simulators share the same recurrent ``GN block + MLP Decoder" architecture but are trained with pipe and bifurcation datasets, respectively. As shown in Fig. \ref{fig:GNNSimulator}, the input feature vectors depicting geometry, parameter settings, boundary conditions and concentration values $c^{t_{k}}$ at $t_k$ are first processed by a series of GN blocks to learn the hidden layer representations that encode both local graph structure and features of nodes. In our GNN simulator, the GN block includes two modules $\phi_e$, $\phi_v$ to update the edge and node features, respectively. $\phi_e$ computes the concentration gradient along the edge and the length of the edge. $\phi_v$ is a multilayer perceptron (MLP) consisting of two hidden layers (with ReLU activations), each layer with the size of 32. The forward computation of our GN block is characterized by Algorithm \ref{alg:GNblock}. After the GN blocks output the latent graph that encodes all the node feature, a MLP is then used to decode the hidden layer representation and output predicted concentration values $c^{t_{k+1}}$ at the next time step $t_{k+1}$. The MLP has three hidden layers (with ReLU activations), followed by a non-activated output layer, each layer with the size of 32.
\begin{algorithm}[H]
    \SetAlgoLined
    \KwIn{Graph, $\mathcal{G} = (\mathcal{V}, \mathcal{E}, x_v)$}
    \For{each edge $e_{ij} \in \mathcal{E}$}{
        Compute edge attributes: $x_{e_{ij}} = \phi_e(x_{v_i}, x_{v_j})$\;
    }
    \For{each node $v_{i} \in \mathcal{V}$}{
    Aggregate edge attributes: $x_{e,v_i} = \sum_j x_{e_{ij}} $\;
    Compute output: $x'_{v_i} = \phi_v(x_{v_i}, x_{e,v_i})$\;
    }
    \KwOut{Graph, $\mathcal{G} = (\mathcal{V}, \mathcal{E}, x'_v)$}
    \caption{GN Block}
    \label{alg:GNblock}
\end{algorithm}

The input graph of the simulator is created by extracting certain nodes from each cross section of a hexahedral mesh following the templates (Supplementary Fig. S1). For each pipe structure, we use the circular plane template (Supplementary Fig. S1C) to extract 17 nodes from each cross section. For each bifurcation structure, we use another template (Supplementary Fig. S1D) to extract 23 nodes from the cross section at the branch point of the bifurcation. We also use the same circular plane template to extract 3 circular cross sections on each branch of the bifurcation. For each node, we collect the geometry information and simulation parameters in an input feature vector. Here, the geometry information of each node is encoded by its coordinates and the radius of the cross section on which the node is located. The nodal concentration value is set to be the target prediction.

To encourage the model to learn the physics underlying the simulation data, we add the residuals of the governing equation (Eq. \ref{equation:transport} in Methods) to the mean squared error (MSE) loss function as
\begin{equation}
    \label{equation:LossFunction_Sim}
    \begin{split}
        \mathcal{L}_{simulator}& =\frac{1}{N}\sum \limits^{N}_{i=1}\left[(c^{P}_{0,i}-c^{G}_{0,i})^2 + (c^{P}_{\pm,i}-c^{G}_{\pm,i})^2\right] + \frac{1}{N}\sum \limits^{N}_{i=1}\left[\dfrac{\partial c_{0,i}}{\partial t}-D \nabla^{2} c_{0,i} + (k_{+}+k_{-})c_{0,i}-k'_{+}c_{+,i}-k'_{-}c_{-,i}\right]^2 \\
        & + \frac{1}{N}\sum \limits^{N}_{i=1}\left[\dfrac{\partial c_{\pm,i}}{\partial t}+\bm{u}_{\pm} \cdot \nabla c_{\pm,i}-k_{\pm}c_{0,i}+k'_{\pm}c_{\pm,i}\right]^2,
    \end{split}
\end{equation}
where $c_{0}$, $c_{+}$ and $c_{-}$ are the spatial concentrations; $D$ is the diffusion coefficient of materials; $\bm{u}_{+}$ and $\bm{u}_{-}$ are velocities of materials; $k_{\pm}$ and $k'_{\pm}$ are filament attachment and detachment rates (See Eq. \ref{equation:transport} in Methods for the detailed physical meaning of the aforementioned variables); $N$ denotes the number of nodes on the graph; and superscripts $P$ and $G$ denote the prediction and the ground truth value on the graph, respectively.

\begin{figure*}[!htbp]
    \centering
    \includegraphics[width = \linewidth]{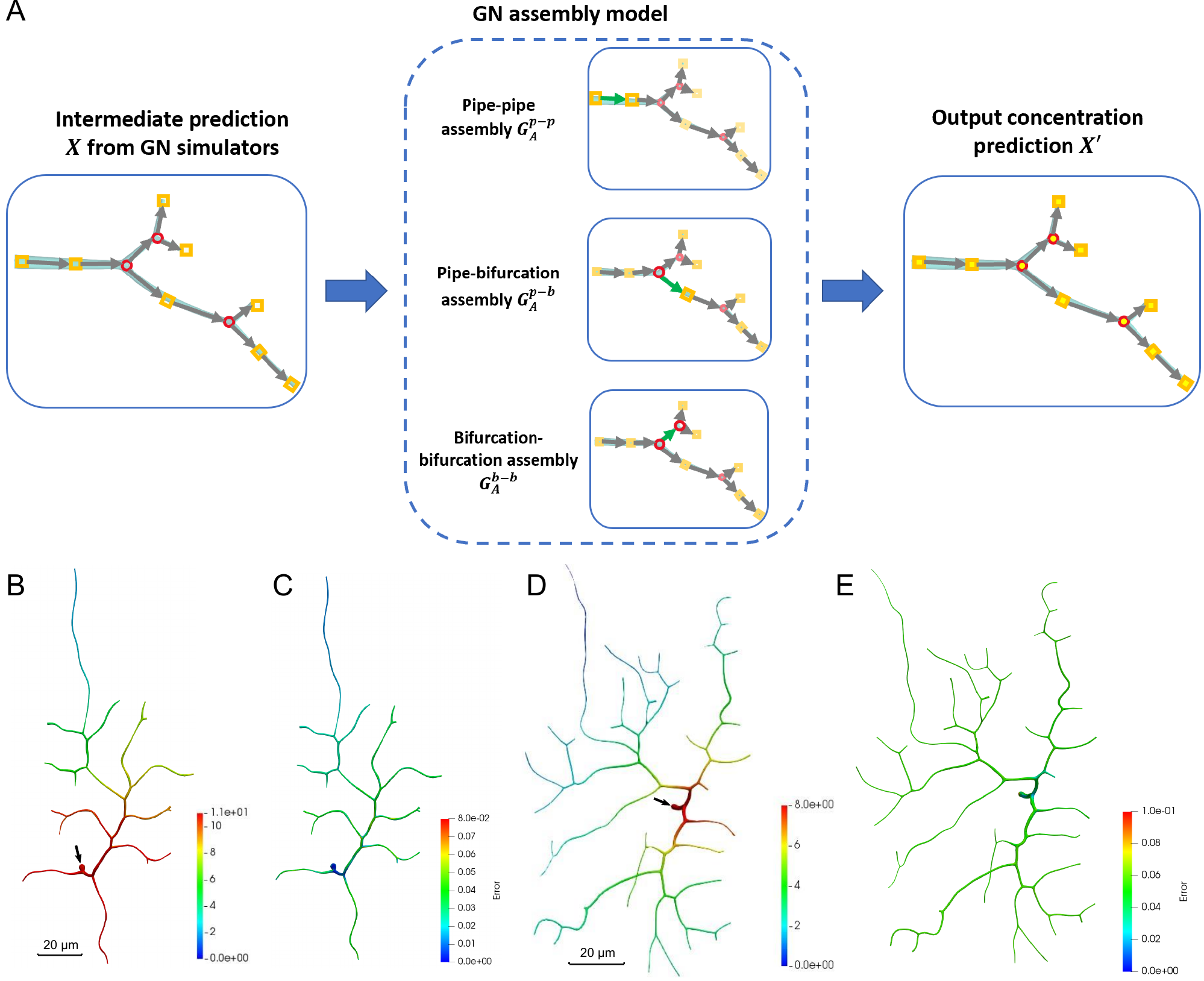}
    \caption{Results of the GNN assembly model. (A) The GNN assembly model takes the intermediate prediction $X$ from individual GNN simulators as input and outputs the final concentration prediction $X'$. The GNN assembly components $G_A^{p-p}$, $G_A^{p-b}$ and $G_A^{b-b}$ adopt the ``message-passing" scheme to aggregate the intermediate prediction from the neighbouring nodes to update the concentration prediction. The green arrows show the message-passing process during the assembly. (B, C) The predicted concentration and the nodal prediction error of NMO\_66731 at steady state ($t=25\;s$). (D, E) The predicted concentration and the nodal prediction error of NMO\_66748 at steady state ($t=31\;s$). In (B) and (D), the black arrows point to the inlet of material. Scalar bar: 20 $\mu m$. Unit for color bars: $mol/\mu m^3$.}
    \label{fig:GNNAssembly}
\end{figure*}

To create the training dataset for two simulators, we first use our IGA solver to run the material transport simulation (see Methods for details) in two complex zebrafish neurons from the NeuroMorpho.Org database \cite{ascoli2007neuromorpho} (NMO\_66731 and NMO\_66748 in Fig. \ref{fig:GNNAssembly}B\&D). Regarding the simulation setting, we use 200 different boundary condition values and set constant parameters in Eq. \ref{equation:transport} and Eq. \ref{equation:NS_bc} as $D=1.0\;\mu m^2/s$, $k_{\pm} = 1.0\;s^{-1}$, $k'_{\pm} = 0.5\;s^{-1}$ and $u_i = 0.1\;\mu m/s$. The time step for each simulation is set to be $0.1\;s$. We simulate until the transport process is steady and then extract the spatiotemporal simulation results of 100 different geometries for each simulator from these two neurons. The nodal feature vectors and nodal concentration values are stored for each geometry to build the training dataset that contains 20,000 samples for each simulator.
After establishing the architecture of two simulators, we train each simulator by randomly selecting 75\% samples as the training data. The Adam optimizer \cite{kingma2014adam} is used to optimize the model with the step learning rate decay ranging from $10^{-3}$ to $10^{-6}$. Our simulators are trained for 200 epochs and we evaluate the performance of the model using the rest 25\% samples as the test dataset. At the end of training, the test loss for the pipe and bifurcation simulators converges to 0.265 and 0.111, respectively (Supplementary Fig. S2).

To demonstrate the performance of our GNN simulators, we select three prediction results of a pipe and a bifurcation from the test dataset and compare with their corresponding IGA simulation results in Fig. \ref{fig:GNNSimulator}. For the pipe simulator, we find the model can accurately capture the boundary conditions and predict the concentration distribution at each time step, which indicates that the GNN simulator manages to learn the time-dependent behaviour of the transport equations. By comparing nodal error at $t = 2.6\;s$ and $t = 3.4\;s$ in Fig. \ref{fig:GNNSimulator}B, we find the error increases along with the front of material propagation through the pipe, which suggests that the pipe simulator is sensitive to the sudden distribution change during the material propagation and needs further improvement. For the bifurcation simulator, the predicted result has higher accuracy around the branch region which suggests that the bifurcation simulator learns to transport the material in correct directions at the branch point. However, the boundary condition is not preserved as well as the pipe simulator performs. The possible reason is that the boundary nodes have less neighbouring nodes for edge feature aggregation compared to the interior nodes and the lack of neighbouring information leads to higher error. Therefore, our bifurcation simulator can be further improved to balance the neighbourhood between the boundary and interior nodes to better preserve the boundary condition.

We perform 4-fold cross validation and the mean relative error (MRE) results for both simulators are shown in the second column of Table \ref{table:CSV}. The well-trained GNN simulators can provide concentration prediction with an error of less than 8\% in average. With the standard deviation less than 1.3\%, the model shows good performance when being generalized to unknown data. To study the impact of the PDE residuals in Eq. \ref{equation:LossFunction_Sim} on the prediction performance, we compare our simulators with the model trained using the standard MSE loss function. We perform cross validation with the same dataset for each comparison and the results are shown in Table \ref{table:CSV}. The comparison shows that the model achieved better accuracy when trained with the ``MSE + PDE residuals'' loss function, which indicates the physics information contained in PDE residuals is learned by the model and improves its performance.
\begin{table}[!htb]
    \vspace{-3mm}
    \caption {MRE comparison between simulators using different loss functions} \label{table:CSV}
    \vspace{-6mm}
    \begin{center}
        \begin{tabular}{||c c c||}
            \hline
            Loss function         & MSE                & MSE                \\
                                  &                    & + PDE residuals    \\
            \hline
            Pipe simulator        & 10.9 $\pm$ 3.45 \% & 6.10 $\pm$ 1.25 \% \\
            Bifurcation simulator & 13.5 $\pm$ 2.47 \% & 7.20 $\pm$ 0.78 \% \\

            \hline
        \end{tabular}
    \end{center}
    \vspace{-9mm}
\end{table}

\subsection*{GNN assembly model}
The objective of the GNN assembly model is to assemble the local prediction from simulators and output an improved continuous concentration distribution on the entire geometry. An overview of the GNN assembly model is shown in Fig. \ref{fig:GNNAssembly}. The assembly model needs to learn the interaction between the simulators. Here the assembly model includes three components to cover all the assembly scenario in a decomposed neurite network: (a) pipe and pipe $G_A^{p-p}$; (b) pipe and bifurcation $G_A^{p-b}$; and (c) bifurcation and bifurcation $G_A^{b-b}$.
During the assembly, the model loops each simulator node on the decomposed neurite network and utilizes the ``message-passing" scheme to gather the predicted results from its neighbouring simulator nodes (green arrows in Fig. \ref{fig:GNNAssembly}A). In particular, the nodal predicted results on the interface between two simulator nodes are collected. Then, all the collected values from the neighbouring simulator nodes are concatenated with the values from the current simulator node and processed by a MLP to improve the prediction. While $G_A^{p-p}$, $G_A^{p-b}$ and $G_A^{b-b}$ have different numbers of input and output values, they share the same MLP architecture with three hidden layers (with ReLU activations), followed by a non-activated output layer, each layer with the size of 32.

To ensure the concentration result is consistent on the interface between two simulators, we add a penalty term to the MSE loss function as
\begin{equation}\label{equation:LossFunction_Assemble}
    \mathcal{L}_{assembly}=\frac{1}{N}\sum \limits^{N}_{i=1}\left\{(c^{P}_{0,i}-c^{G}_{0,i})^2 + (c^{P}_{\pm,i}-c^{G}_{\pm,i})^2\right\} + \alpha \frac{1}{M}\sum \limits^{M}_{j=1}\left\{(c^{s_1, interface}_{0,j}-c^{s_2, interface}_{0,j})^2 \right. \left. + (c^{s_1, interface}_{\pm,j}-c^{s_2, interface}_{\pm,j})^2\right\},
\end{equation}
where $N$ denotes the number of nodes on two assembled simulators, $M$ denotes the number of nodes on the interface, superscript $P$ and $G$ denote the prediction and the ground truth value on the graph, respectively. Superscripts $s_1$ and $s_2$ denote the prediction value from the first and second simulators, respectively. $\alpha$ is the penalty strength which is set to be 10 in this study.

We use the same IGA simulation results in two complex geometries of zebrafish neurons (NMO\_66731 and NMO\_66748 in Fig. \ref{fig:GNNAssembly}B\&D) to create training dataset for the assembly model. Based on the graph representation of these two trees, we follow three basic assembly structures and extract 30 different geometries for each type of assembly.
The simulation results of these geometries are output and create 6,000 samples for each assembly structure.
The model is trained using 75\% samples as the training dataset and we evaluate the performance of the model using the rest 25\% samples as the test dataset. At the end of training, the test loss for the GNN assembly model converges to 0.1492 (Supplementary Fig. S3). To test the performance of the assembly model, we use these two complex geometries and compare the concentration prediction at steady state (Fig. \ref{fig:GNNAssembly}B-E). The prediction MREs are 6.7\% for NMO\_66731 and 7.3\% for NMO\_66748, respectively. We find that the assembly model manages to generate a continuous global concentration distribution for each geometry. The comparison of predicted concentration between nodal error for each geometry also shows that the high concentration regions are predicted accurately. This indicates that our GNN model can capture the locations that are easier to develop transport disruption due to material accumulation.
\begin{figure*}[!htb]
    \centering
    \includegraphics[width = \linewidth]{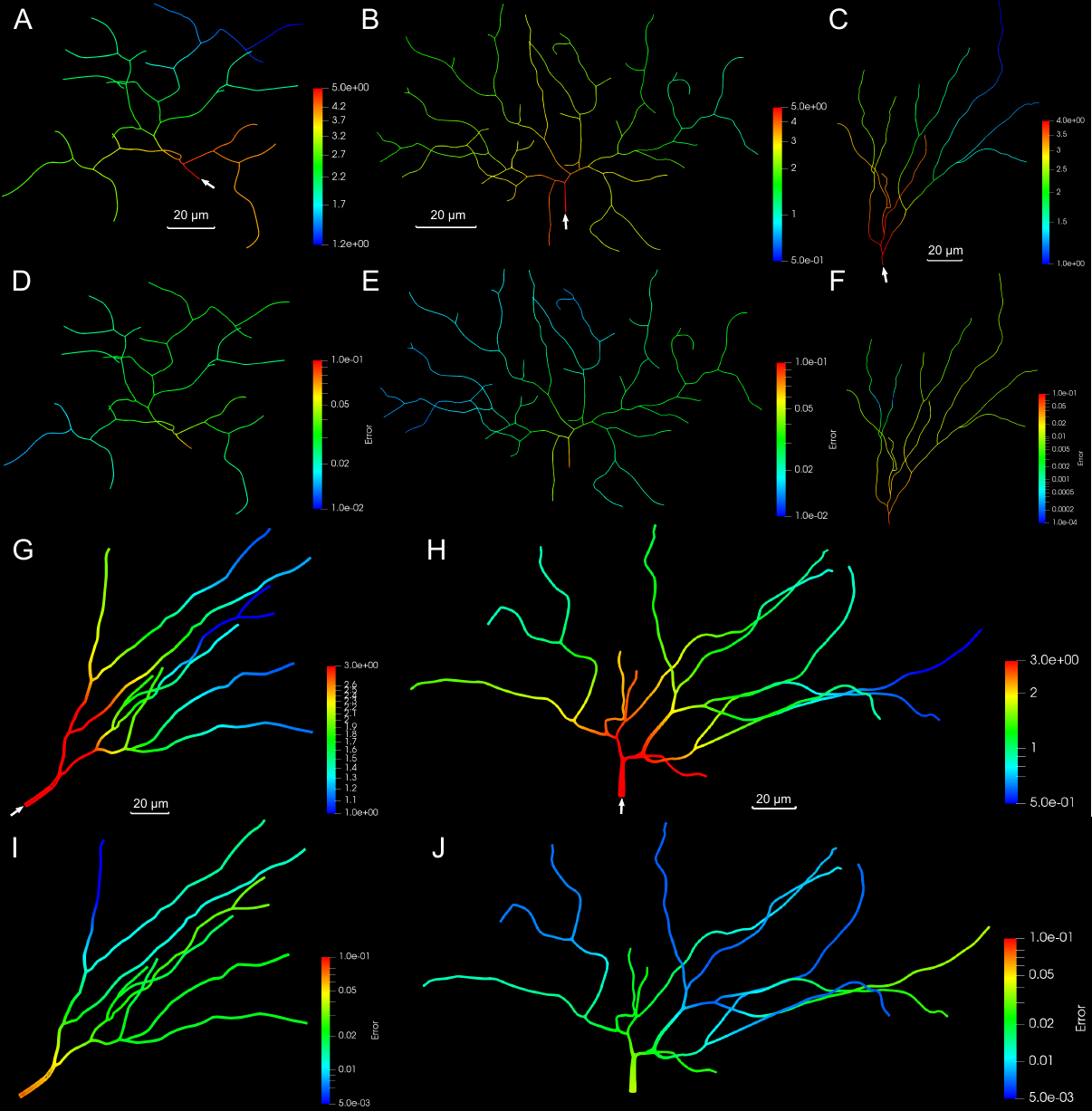}
    \caption{The concentration prediction of material transport in the neurite network geometry of NMO\_06846, NMO\_06840, NMO\_112145, NMO\_32235 and NMO\_32280. (A-C, G, H) The predicted concentration results of steady state at $t=15\;s$, $26\;s$, $19\;s$, $28\;s$ and $32\;s$, respectively. The white arrows point to the inlet of material. Scalar bar: 20 $\mu m$. Unit for color bars: $mol/\mu m^3$. (D-F, I, J) The nodal errors between the ground truth and predicted concentration. Logarithmic scale is used to highlight the distribution pattern.}
    \label{fig:Result_ComplexNeuriteNet3}
    \vspace{-5mm}
\end{figure*}
\begin{figure*}[!htb]
    \centering
    \includegraphics[width = \linewidth]{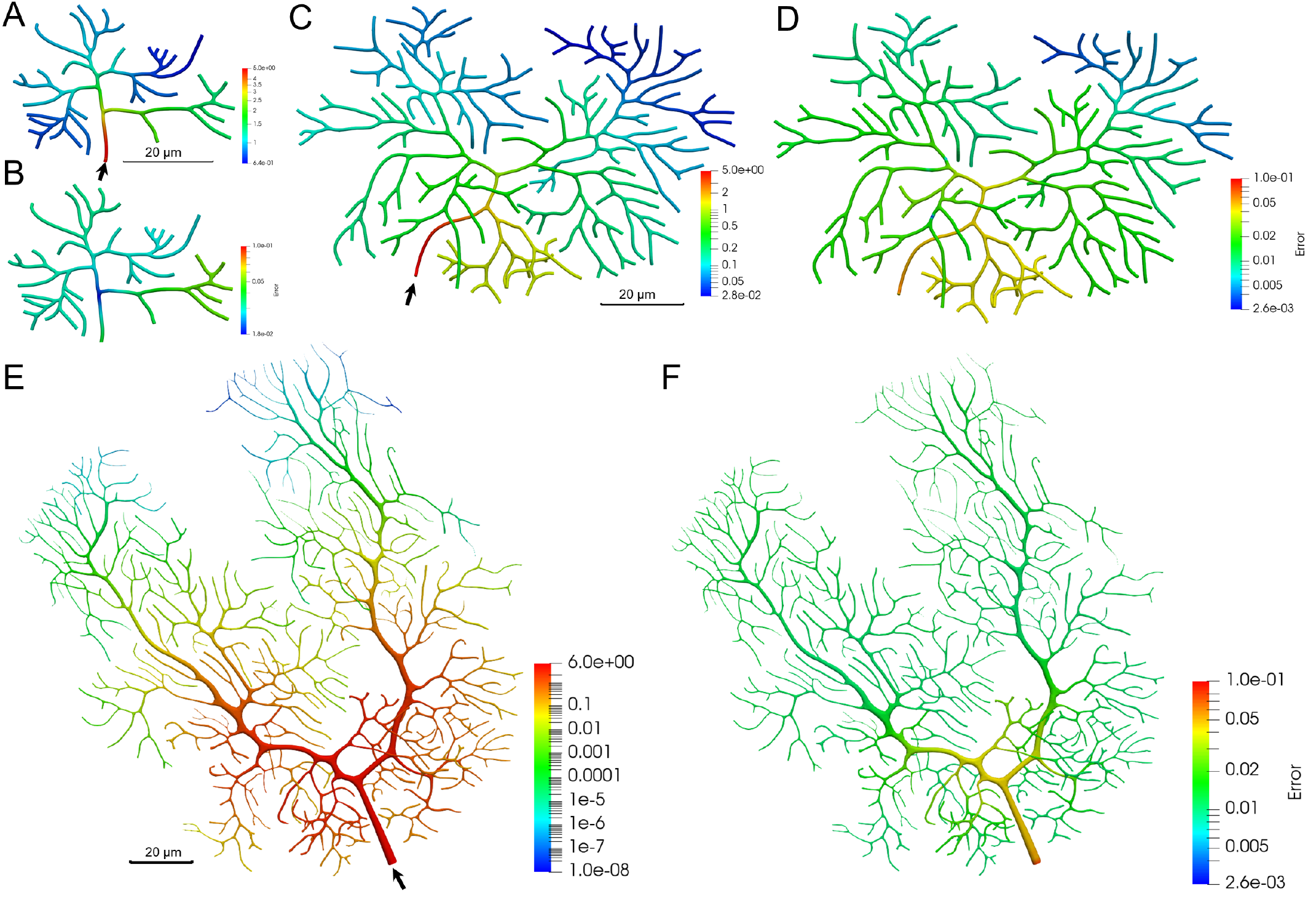}
    \caption{The concentration prediction of material transport in the neurite network geometry of NMO\_54504, NMO\_54499 and NMO\_00865. (A, C, E) The predicted concentration results of steady state at $t=12\;s$, $41\;s$, and $63\;s$, respectively. The black arrows point to the inlet of material. Scalar bar: 20 $\mu m$. Unit for color bars: $mol/\mu m^3$. (B, D, F) The nodal errors between the ground truth and predicted concentration. Logarithmic scale is used to highlight the distribution pattern.}
    \label{fig:Result_ComplexNeuriteNet4}
    \vspace{-5mm}
\end{figure*}

\subsection*{Results for complex neurite networks}
After the simulators and the assembly model in the GNN framework are well trained, we use our GNN model to predict the concentration distribution in several complex neurite networks and compare with the simulation results from our IGA solver. All complex neurite networks selected from the NeuroMorpho.Org database are shown in Figs. \ref{fig:Result_ComplexNeuriteNet3} and \ref{fig:Result_ComplexNeuriteNet4}.
Since the GNN model is trained with the simulation data in zebrafish neurons, we pick another two zebrafish neurons (Fig. \ref{fig:Result_ComplexNeuriteNet3}A\&B) to test the model performance in the neurons from the same species. We also pick another six mouse neurons (Fig. \ref{fig:Result_ComplexNeuriteNet3}C, \ref{fig:Result_ComplexNeuriteNet3}G, \ref{fig:Result_ComplexNeuriteNet3}H and \ref{fig:Result_ComplexNeuriteNet4}) from a different species to validate the model.
We choose neurons with the number of bifurcations ranging from 9 to 356.
For each neuron, we first run IGA simulation to get the ground truth concentration results and then compare with the predicted concentration obtained from our GNN model. Fig. \ref{fig:Result_ComplexNeuriteNet3} shows the prediction results for the neurons with longer branches and Fig. \ref{fig:Result_ComplexNeuriteNet4}, shows neurons with more bifurcations and shorter branches. We observe that our GNN model predicts the distribution well with relative larger error around the inlet regions with high concentration. In addition, the model performance gets worse in longer branches or regions with high density of bifurcations. The possible reason is that the increasing complexity of the geometry leads to a larger graph representation which affects the prediction accuracy due to the complicated assembling process on more simulators.
\begin{table*}
    \vspace{-3mm}
    \caption {Statistics of all tested complex neurite networks}
    \vspace{-3mm}
    \resizebox{\textwidth}{!}{
        \begin{tabular}{||c c c c c c c c||}
            \hline
            \multirow{2}{*}{Species}   & \multirow{2}{*}{Model name}                             & Mesh                   & Bifurcation & IGA computation      & GNN prediction & Speedup ratio & GNN prediction \\
                                       &                                                         & (vertices, elements)   & number      & (nodes, time (mins)) & time (mins)    & (IGA vs GNN)  & MRE            \\
            \hline
            \multirow{4}{*}{Zebrafish} & NMO\_66731 (Fig. \ref{fig:GNNAssembly}B)                & (127,221, 112,500)     & 15          & (8, 468)             & 1.6            & 293           & 6.7\%          \\
                                       & NMO\_66748 (Fig. \ref{fig:GNNAssembly}D)                & (282,150, 249,660)     & 35          & (10, 672)            & 3.9            & 172           & 7.3\%          \\
                                       & NMO\_06846 (Fig. \ref{fig:Result_ComplexNeuriteNet3}A)  & (116,943, 101,880)     & 20          & (8, 413)             & 2.1            & 197           & 7.4\%          \\
                                       & NMO\_06840 (Fig. \ref{fig:Result_ComplexNeuriteNet3}B)  & (280,434, 248,040)     & 35          & (10, 705)            & 4.4            & 160           & 7.6\%          \\
            \hline
            \multirow{6}{*}{Mouse}     & NMO\_112145 (Fig. \ref{fig:Result_ComplexNeuriteNet3}C) & (110,985, 98,460)      & 9           & (8, 350)             & 1.3            & 269           & 8.2\%          \\
                                       & NMO\_32235 (Fig. \ref{fig:Result_ComplexNeuriteNet3}H)  & (96,714, 85,680)       & 9           & (6, 320)             & 1.1            & 291           & 8.3\%          \\
                                       & NMO\_32280 (Fig. \ref{fig:Result_ComplexNeuriteNet3}I)  & (131,967, 117,000)     & 12          & (8, 493)             & 1.5            & 329           & 8.1\%          \\
                                       & NMO\_54504 (Fig. \ref{fig:Result_ComplexNeuriteNet4}A)  & (116,943, 101,880)     & 32          & (8, 436)             & 3.3            & 132           & 8.3\%          \\
                                       & NMO\_54499 (Fig. \ref{fig:Result_ComplexNeuriteNet4}C)  & (524,871 459,360)      & 127         & (20, 759)            & 6.1            & 124           & 8.9\%          \\
                                       & NMO\_00865 (Fig. \ref{fig:Result_ComplexNeuriteNet4}E)  & (1,350,864, 1,179,900) & 356         & (40, 908)            & 7.2            & 126           & 9.2\%          \\
            \hline
        \end{tabular}
    }
    \begin{tablenotes}
        \small \item Note: Each node in Bridges has 28 cores.
    \end{tablenotes}
    \label{table:stat_complex}
    \vspace{-6mm}
\end{table*}
The detailed geometry information, computation statistics and error comparison are summarized in Table \ref{table:stat_complex}. By comparing the prediction MRE of zebrafish neurons (top four rows in Table \ref{table:stat_complex}), we find the MRE only increases by around 0.9\% for all the tested zebrafish neurons, which validates the applicability of the model among neurons from the same species. In addition, we find that the average prediction MREs of zebrafish and mouse neurons are comparable with 7.25\% and 8.5\%, respectively, indicating that the trained model can also work for neurons from other species.

To evaluate the computation performance, we define a speedup ratio as the ratio between the IGA computation time and the GNN prediction time.
We run the code and measure the computation time on the XSEDE (Extreme Science and Engineering Discovery Environment) supercomputer Bridges \cite{towns2014xsede} in the Pittsburgh Supercomputer Center. For each geometry, we perform the IGA simulation through parallel computing on CPU and perform GNN prediction on GPU.
The speedup ratio for each geometry is shown in Table \ref{table:stat_complex} and we find that our GNN model can achieve up to 330 times faster compared to IGA simulation. We also observe that the speedup ratio decreases when the neuron geometry becomes complicated with more bifurcations. The ratio converges to around 120 when the bifurcation number reaches 356, which is still a significant improvement by reducing computation time from hours to minutes.

\section*{Discussion}
\label{section:Discussion}
In this paper, we present a GNN-based DL model for predicting the material transport process in complex geometry of neurons. Our strategy utilizes a graph representation of neurons to overcome the difficulty in handling different neuron geometries with the DL model. The underlying assumption of our approach is that the material concentration distribution can be predicted locally in simple geometries and then be assembled following the graph representation to restore the concentration distribution in any complex geometry. Two GNN models are developed to validate our assumption. The first GNN-based model serves as the simulator to predict the dynamic concentration results locally on the mesh of two basic structures: pipe and bifurcation. We adopt GNN to directly handle the IGA simulation data stored in the unstructured mesh. Given input boundary conditions and geometry information of a pipe or bifurcation, the well trained simulators are able to provide high accuracy results (MRE $<7\%$). The second GNN-based assembly model then collects all the local predictions and follows the graph representation of the neurite network to update the global prediction. As shown in Figs. \ref{fig:Result_ComplexNeuriteNet3}-\ref{fig:Result_ComplexNeuriteNet4}, the model is evaluated on complex neurons from mouse and zebrafish and shows its applicability with consistent performance in different neuron geometries. In particular, the model is capable of providing the spatiotemporal concentration prediction with MRE $<10\%$ and over 100 times faster than the IGA simulation. Furthermore, the accurate distribution prediction can help us determine high concentration regions, which is essential to infer possible locations to develop transport disruption.

Our study shows that the complex and diverse geometry of neurons has major impact on the material concentration distribution and further affects the prediction accuracy of our GNN model. As shown in Figs. \ref{fig:Result_ComplexNeuriteNet3}-\ref{fig:Result_ComplexNeuriteNet4}, the radius of most complex neurons is larger around the inlet region and decreases downstream, which contributes to material accumulation at the inlet region. The increase of bifurcations can also aggravate the accumulation when there is sharp decrease of radius in the downstream branches (Fig. \ref{fig:Result_ComplexNeuriteNet4}E). These geometry features contribute to the complicated concentration distribution and thus bring challenges to improve the performance of our GNN model. One challenge we have for the most complex neuron NMO\_00865 is the MRE gets worse to 12.5\% when it was initially tested using the model trained with zebrafish neuron dataset. We plot the nodal error and find the error is higher at regions with high curvature or sharp radius change (Supplementary Fig. S5). The possible reason is the GNN model lacks the knowledge of these scenarios since they are not common in the training dataset obtained from zebrafish neurons. To address this issue, we extract 20 geometries for each scenario from NMO\_00865 and include them into the training dataset. We adopt the transfer learning method \cite{pan2009survey} to reuse the pre-trained GNN model as the starting point and obtain an improved model by training with the new dataset. The new model achieves the MRE of 9.2\% on NMO\_00865 which is comparable to the other testing neurons. This indicates that our GNN model can be further optimized with transfer learning on a better training dataset and an optimal dataset with a variety of geometries considered is quite essential to improve the applicability of the model.

The integration of the governing equations with the GNN model also plays a critical role in guiding the GNN model to learn the underlying physics behind the simulation data. By including the PDE residuals as the physics loss in the loss function of the GNN simulator (Eq. \ref{equation:LossFunction_Sim}), a superior performance is achieved over the model trained with the standard MSE loss function. The interface loss is also considered in the loss function of the GNN assembly model (Eq. \ref{equation:LossFunction_Assemble}) to minimize the noncontinuous concentration gap between the assembled geometries, which enables a continuous global concentration prediction. The results indicate that the physics-based loss function serves as an explainable component of the GNN model and teaches the model to utilize the simulation data more efficiently.

Our method, directly learning the transport mechanism from numerical simulation data in mesh format, makes it flexible to design data-driven DL models and further explore the value of simulation data. The model can also be extended to a general GNN framework for learning other PDE models in any complex geometry. Specifically, we can modify the PDE residuals in the simulator loss function to create a physics-guided data-driven DL model for any PDE solver. Our study also has its limitations, which we will address in our future work. In the current model, we only consider different geometries and boundary conditions as input features, while fixing the simulation parameters. In addition, there are still many more neurite networks with different topologies that are not considered in our dataset. To further improve the performance of our GNN model, we will optimize the training dataset by including different simulation parameter settings and different geometries to generalize its application in neurons of broader morphology.
We will also study how to combine GNN with physics-informed neural network so that the geometry information of the data can be fully encoded and utilized in the DL framework.
Despite these limitations, our GNN model efficiently and accurately predicts the dynamic concentration distribution within complex neurite networks and provides a powerful tool for further study in this field.

\section*{Methods}

\subsection*{IGA-based material transport simulation on complex neuron geometries}
In our previous study, we developed an IGA-based solver to perform dynamic material transport simulation in large and complex neurite networks \cite{li2019isogeometric}.
IGA is an advanced finite element technique that differs from conventional finite element analysis (FEA) for its direct integration of geometrical modeling with numerical solution. With the same smooth spline basis functions \cite{piegl2012nurbs} utilized as the basis for both geometrical modeling and numerical solution, IGA can accurately represent a wide range of complex geometry with high-order continuity while offering superior performance over FEA in numerical accuracy and robustness \cite{HUGHES20054135}. Due to its many performance advantages, IGA has been adopted in a wide variety of research areas, such as linear elasticity \cite{dimitri2014isogeometric}, shell analysis \cite{benson2010isogeometric, casquero2020seamless, casquero2017arbitrary}, cardiovascular modeling \cite{zhang2007patient,zhang2012atlas,urick2019review,yu2020anatomically, zhang2016geometric} and fluid-structure interaction \cite{bazilevs2006isogeometric, casquero2016hybrid, casquero2018non}, as well as collocation \cite{anitescu2015isogeometric, casquero2016isogeometric}. Truncated T-splines \cite{wei2017truncated1, wei2017truncated2} were developed to facilitate local refinement over unstructured quadrilateral and hexahedral meshes. Blended B-splines \cite{wei2018blended} and Catmull-Clark subdivision basis functions \cite{li2019hybrid} were investigated to enable improved or even optimal convergence rates for IGA.
Our geometric modeling and IGA simulation pipeline \cite{li2019isogeometric} is shown in Supplementary Fig. S4. To reconstruct the geometry of different neurite networks for the simulation, we first obtain their morphologies stored in the SWC format from the NeuroMorpho database \cite{ascoli2007neuromorpho}. With the skeleton and diameter information provided in the SWC files, we adopt the skeleton-sweeping method \cite{zhang2007patient} to generate the hexahedral control mesh of neurite network, and then build trivariate truncated hierarchical B-splines over it. Regarding the governing equations of the simulation, we generalize the 1D motor-assisted transport model \cite{smith2001models} to 3D geometry to accurately account for the actual morphology of the neurite.
The model is described as a group of ``reaction-diffusion-transport" equations, we have
\begin{equation}\label{equation:transport}
    \left\{
    \begin{array}{lr}
        \dfrac{\partial c_{0}}{\partial t}-D \nabla^{2} c_{0}=-(k_{+}+k_{-})c_{0}+k'_{+}c_{+}+k'_{-}c_{-}     & \text{in $\Omega$},\vspace{1.5mm} \\
        \dfrac{\partial c_{\pm}}{\partial t}+\bm{u}_{\pm} \cdot \nabla c_{\pm}=k_{\pm}c_{0}-k'_{\pm}c_{\pm}\  & \text{in $\Omega$},               \\
        c_{0}=c,\indent c_{+}=\lambda c                                                                       & \text{at incoming end},           \\
        c_{0}=\tilde{c},\indent c_{-}=\tilde{\lambda}\tilde{c}                                                & \text{at outgoing end},
    \end{array}
    \right.
\end{equation}
where the open set $\Omega \subset \mathbb{R}^3$ represents the internal space of the single neurite; $c_{0}$, $c_{+}$ and $c_{-}$ are the spatial concentrations of free, incoming (relative to the cell body; retrograde), and outgoing (anterograde) materials, respectively; $D$ is the diffusion coefficient of free materials; $\bm{u}_{+}$ and $\bm{u}_{-}$ are velocities of incoming and outgoing materials, respectively; $k_{\pm}$ and $k'_{\pm}$ are rates of cytoskeletal filament attachment and detachment of incoming and outgoing materials, respectively; and $\lambda$, $\tilde{\lambda}$ represent the degree of filament attachment at both ends and are also referred to as the ``degree of loading" \cite{smith2001models}. Regarding the boundary condition, we assume stable concentrations of free and incoming materials at both the incoming end and the outgoing end and set constant values to $\lambda$ and $\tilde{\lambda}$. In this study, we assume the filament system is unipolar that leads to an unidirectional material transport process and ignore $c_{-}$, $\bm{u}_-$, $k_-$, $k'_-$ terms in Eq. \ref{equation:transport}.

To obtain a physically realistic transport process in 3D, we assume that the flow of transport is incompressible and solve the steady-state Navier-Stokes equation to derive a physically realistic velocity field inside the single neurite
\begin{equation}\label{equation:NSequation}
    \begin{cases}
        \nabla\cdot\bm{u}=0                                                & \text{in $\Omega$}, \\
        \nabla\cdot(\bm{u}\otimes\bm{u})+\nabla p=\nu \Delta \bm{u}+\bm{f} & \text{in $\Omega$},
    \end{cases}
\end{equation}
where the open set $\Omega \subset \mathbb{R}^3$ represents the incompressible fluid domain, $\bm{u}$ is the flow velocity, $p$ is the pressure, $\bm{f}$ is the given body force per unit volume, $\nu$ is the kinematic viscosity, and $\otimes$ denotes the tensor product. Regarding boundary conditions, we impose non-slip condition at the neurite wall and apply a parabolic profile inlet velocity for each point on the circular cross section as
\begin{equation}
    u(r) = u_i(1-(r/R)^2),
    \label{equation:NS_bc}
\end{equation}
where $u_i$ is the inlet transport velocity defined in our material transport model, $r$ is the distance from the center of the circular cross section to the point, and $R$ is the radius of the circular cross section. The direction of the velocity is perpendicular to the inlet cross section.
We utilize IGA to solve Eq. \ref{equation:NSequation} and get the velocity field $\bm{u}$. Based on the unidirectional transport assumption, the velocity field is used as $\bm{u}_+ = \bm{u}$ to solve Eq. \ref{equation:transport}.
As a result, we obtain the material concentration at every time step stored in the truncated hierarchical B-spline of the neurite network. These simulation results are then collected and processed as the training data for this study.

\subsection*{A review of GNNs}
GNN is a machine learning technique that was first proposed to generalize the existing neural networks to operate on the graph domain \cite{scarselli2008graph}. It has been widely used to perform graph analysis in social networks \cite{hamilton2017inductive,zhang2018link}, traffic networks \cite{geng2019spatiotemporal} and physics \cite{battaglia2016interaction, santoro2017simple}. In the following, we explain the basic idea of GNN by using the original GNN framework \cite{scarselli2008graph}.

A graph $\mathcal{G}$ is a pair $(\mathcal{V}, \mathcal{E})$, where $\mathcal{V}$ is the set of nodes and $\mathcal{E}$ is the set of edges. Given the node features $\textbf{X}_{\mathcal{V}}$, the edge features $\textbf{X}_{\mathcal{E}}$ and outputs $\textbf{O}$, the GNN is trained to learn the embedding state $\textbf{S}$ to establish the global mapping between all the features and outputs. Since each node is naturally defined by its features and the related nodes in a graph, the nodal embedding state $\textbf{s}_v$ and the nodal output $\textbf{o}_v $ can be produced locally as
\begin{align}
    \textbf{s}_v & = f_t(\textbf{x}_v, \textbf{x}_\epsilon, \textbf{s}_{ne[v]}, \textbf{x}_{ne[v]}), \\
    \textbf{o}_v & = f_o(\textbf{s}_v,\textbf{x}_v),
\end{align}
where $\textbf{x}_v, \textbf{x}_e, \textbf{s}_{ne[v]}, \textbf{x}_{ne[v]}$ are the features of node $v$, the features of its edges, the embedding states and the features of the nodes in the neighborhood of $v$, respectively. $f_t$ is the \textit{local transition function} and $f_o$ is the \textit{local output function}. Both $f_t$ and $f_o$ are the parametric functions including the parameters to be trained and shared among all nodes.

Let $F_t$ (the \textit{global transition function}) and $F_o$ (the \textit{global output function}) be stacked versions of $f_t$ and $f_o$ for all nodes in a graph, respectively. Then, we get a compact form of the global mapping:
\begin{align}
    \textbf{S} & = F_t(\textbf{S}, \textbf{X}_{\mathcal{V}},\textbf{X}_{\mathcal{E}}), \\
    \textbf{O} & = F_o(\textbf{S},\textbf{X}_{\mathcal{V}}).
\end{align}
Based on Banach's fixed point theorem \cite{khamsi2011introduction}, GNN uses the following iterative scheme to compute the embedding states:
\begin{equation}
    \textbf{S}^{k+1} = F_t(\textbf{S}^{k}, \textbf{X}_{\mathcal{V}},\textbf{X}_{\mathcal{E}}),
\end{equation}
where $\textbf{S}^{k}$ is the $k$-th iteration of $\textbf{S}$.
The training of the aforementioned GNN model is straightforward. With the target output $\tilde{\textbf{o}}_v$ for the supervision, the loss function can be defined as
\begin{equation}
    loss = \sum_{i=1}^{p}||\tilde{\textbf{o}}_{v,i} - \textbf{o}_{v,i}||,
\end{equation}
where $p$ is the number of supervised nodes. Then the gradient-descent strategy is used to update and learn the parameters in $f_t$ and $f_o$.

The original GNN framework has the limitation that they can only handle the nodes embedded in fixed graphs and have to be re-trained whenever the graph structure changes. To address this limitation, GraphSAGE \cite{hamilton2017inductive} was proposed to generalize the GNN algorithms to learn the nodes embedded in dynamic graphs. The key idea of GraphSAGE is to learn how to aggregate feature information from a node’s local neighborhood. When the aggregator weights are learned, the embedding of an unseen node can be generated from its features and neighborhood. To further generalize the concept of GNN, several different GNN frameworks were proposed to integrate different DL models into one framework, such as the message passing neural network (MPNN) \cite{gilmer2017neural}, the non-local neural network (NLNN) \cite{wang2018non} and the graph network (GN) \cite{battaglia2018relational}. In our study, the extensive geometry of neurite networks contributes to different mesh structures for prediction. In particular, different lengths of the pipe lead to different numbers of cross sections along the pipe skeleton. Therefore, we implement GN in our GNN model to ensure the model is suitable for any neuron geometry.

\subsection*{Model evaluation}
Two performance metrics were used to evaluate the accuracy of the predicted concentration distributions from the our algorithm: mean absolute error (MAE) and mean relative error (MRE). For each predicted result, the MAE is defined by
\begin{equation}
    MAE=\sqrt{\sum \limits^{N}_{i=1} \frac{1}{N}(c^{P}_{i}-c^{G}_{i})^2},
    \label{equation:MAE}
\end{equation}
where $N$ denotes the number of elements in the output, $c^{P}_{i}$ and $c^{G}_{i}$ denote the predicted and ground truth concentration values of the $i^{th}$ node in a given mesh, respectively. For each predicted result, the MRE is defined by
\begin{equation}
    MRE=\frac{MAE}{\max \left|\left|c^G\right|\right| - \min \left|\left|c^G\right|\right|} \times 100\%,
    \label{equation:MRE}
\end{equation}
where $\max \left|\left|c^G\right|\right|$ and $\min \left|\left|c^G\right|\right|$ denote the maximum and minimum nodal concentration values from the ground truth result, respectively.

\subsection*{Code and data availability.} The source code for our model and all input data are available for download from a public software repository located at \url{https://github.com/truthlive/NeuronTransportLearning}. All data generated during this study can be reconstructed by running the source code.

\bibliography{reference}

\section*{Acknowledgements}

The authors acknowledge the support of NSF grants CMMI-1953323 and CBET-1804929. This work used the Extreme Science and Engineering Discovery Environment (XSEDE), which is supported by the NSF grant ACI-1548562. Specifically, it used the Bridges system, which is supported by the NSF grant ACI-1445606, at the Pittsburgh Supercomputing Center (PSC).

\section*{Author contributions statement}
A.L. designed the model and carried out the calculations and analysis with the instruction from Y.J.Z.. A.B.F. provided advice on the selection of machine learning methods. All authors contributed to the writing, discussions and revisions of the manuscript.

\section*{Additional information}
\subsection*{Competing interests.} The authors declare no competing interests.


  
  
  
  

  \newpage
  \setcounter{figure}{0}
\makeatletter
\renewcommand{\thefigure}{S\arabic{figure}}
  
  \section*{Supplementary Figures}
  
  \begin{figure*}[!ht]
    \includegraphics[width=1.0\textwidth]{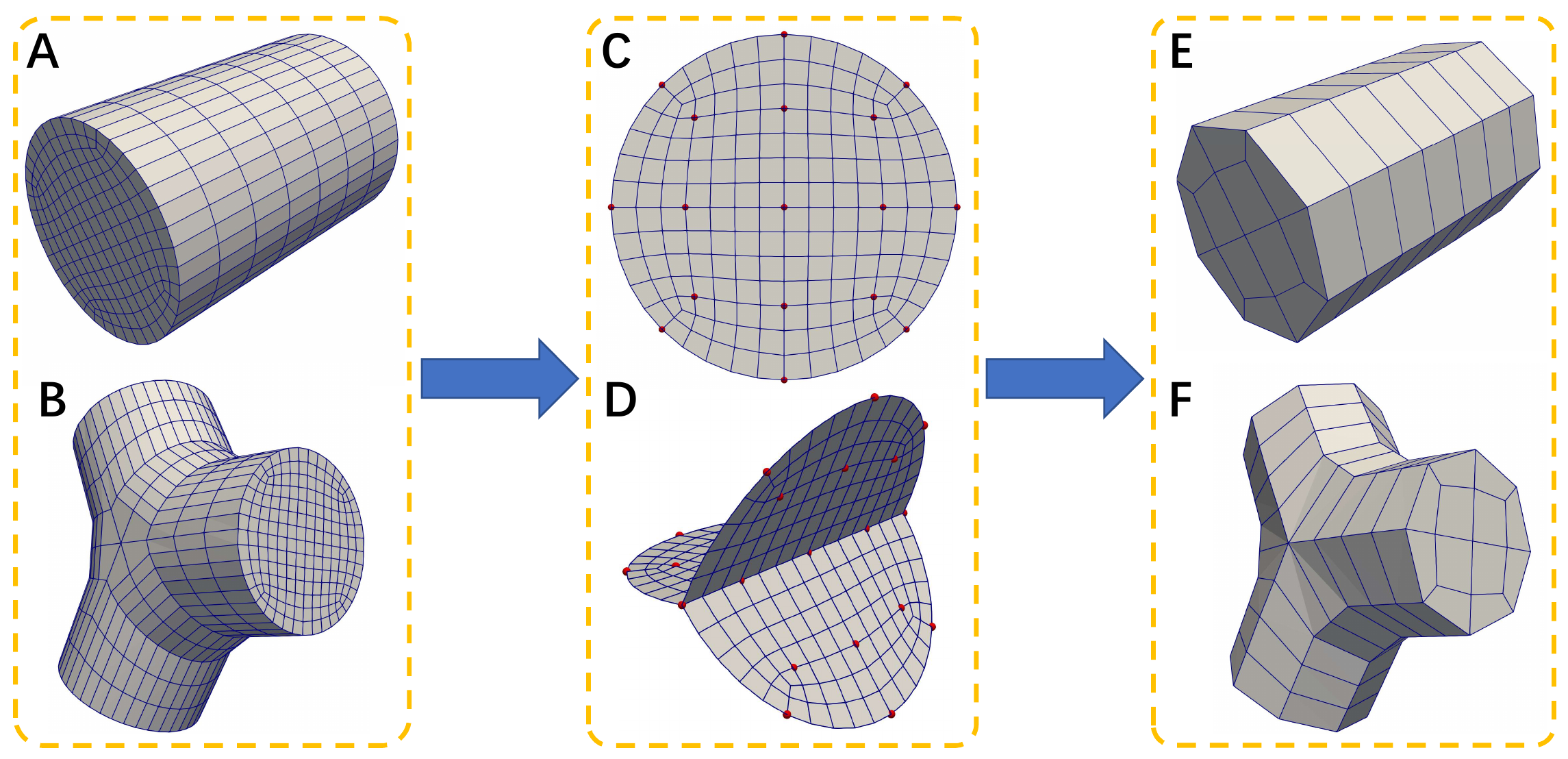}
    \caption{The graph extraction of the pipe and bifurcation structures. (A, B) The input control meshes of the pipe and the bifurcation. (C) The extraction template for circular cross sections with 17 extracted nodes labeled in red. (D) The extraction template for branch cross sections with 23 extracted nodes labeled in red. (E, F) The extracted graphs of the pipe and the bifurcation.}
    \label{figS:ExtractSimulator}
  \end{figure*}
  
  \begin{figure*}[!ht]
    \centering
    \includegraphics[width = 0.9\linewidth]{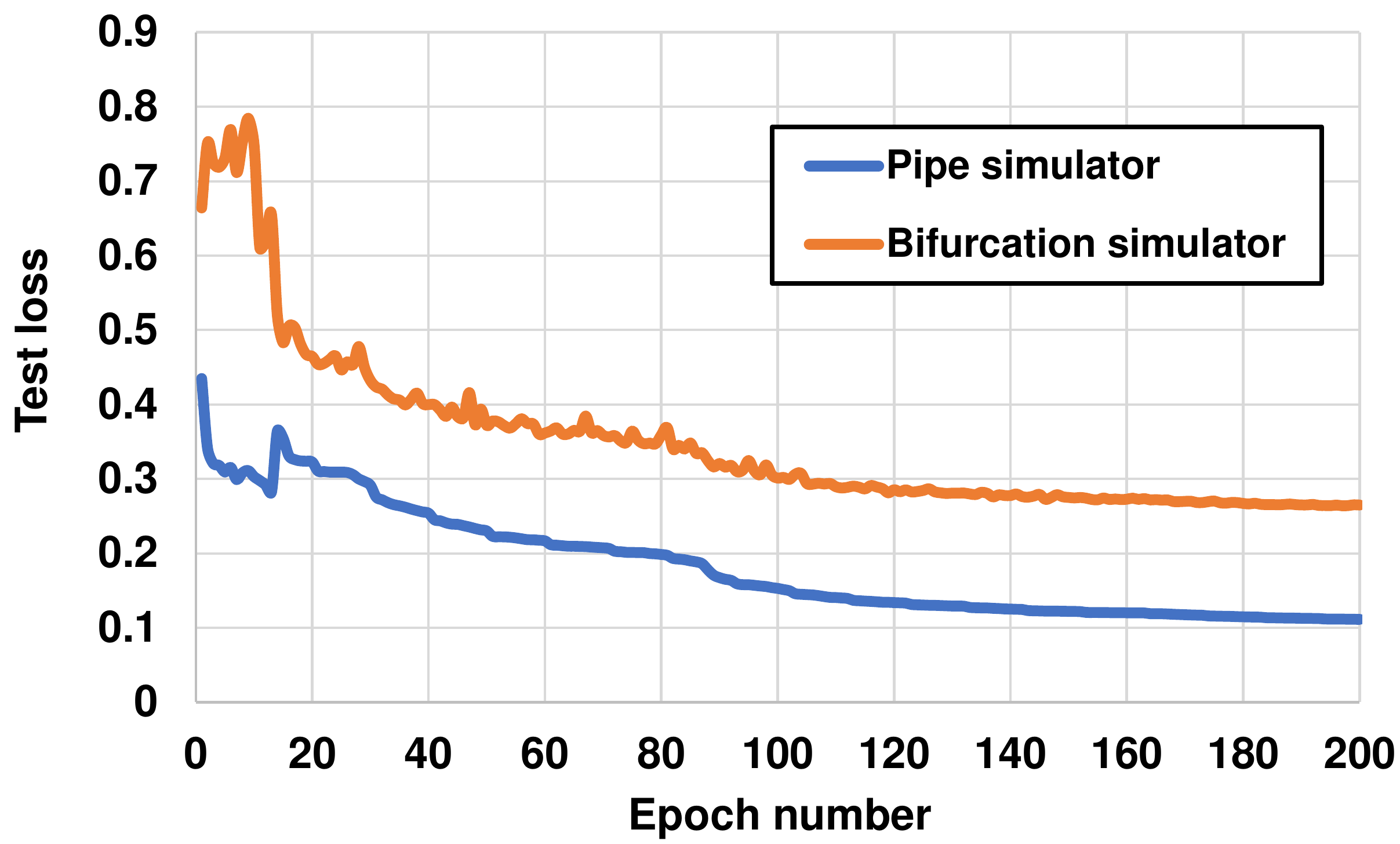}
    \caption{The test loss vs epoch curves for two GNN simulators.}
    \label{figS:Result_EpochLoss_Simulator}
  \end{figure*}
  
  \begin{figure*}[!ht]
    \centering
    \includegraphics[width = 0.9\linewidth]{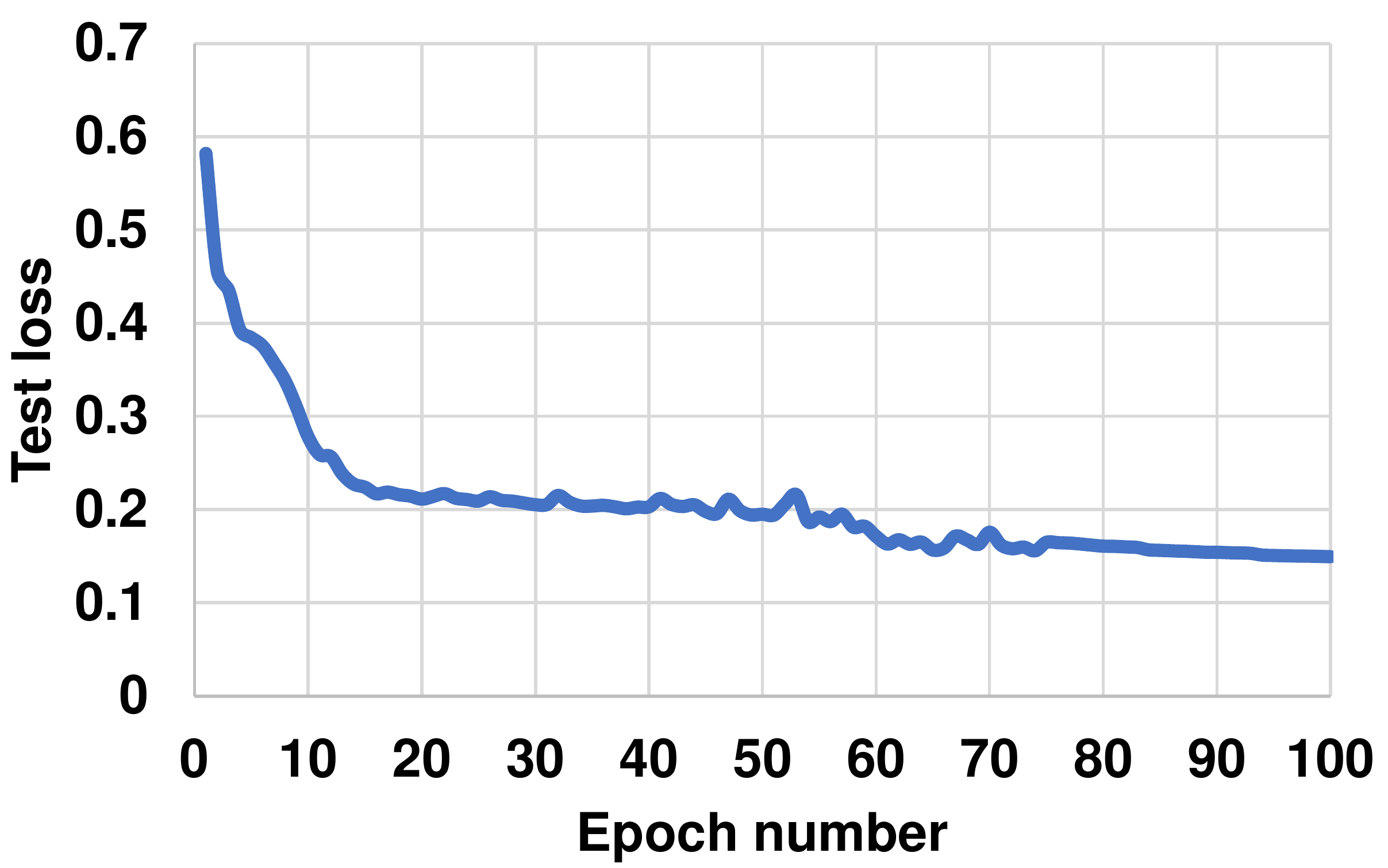}
    \caption{The test loss vs epoch curve for the GNN assembly model.}
    \label{figS:Result_EpochLoss_Assembly}
  \end{figure*}
  
  \begin{figure*}[!htb]
    \centering
    \includegraphics[width = \linewidth]{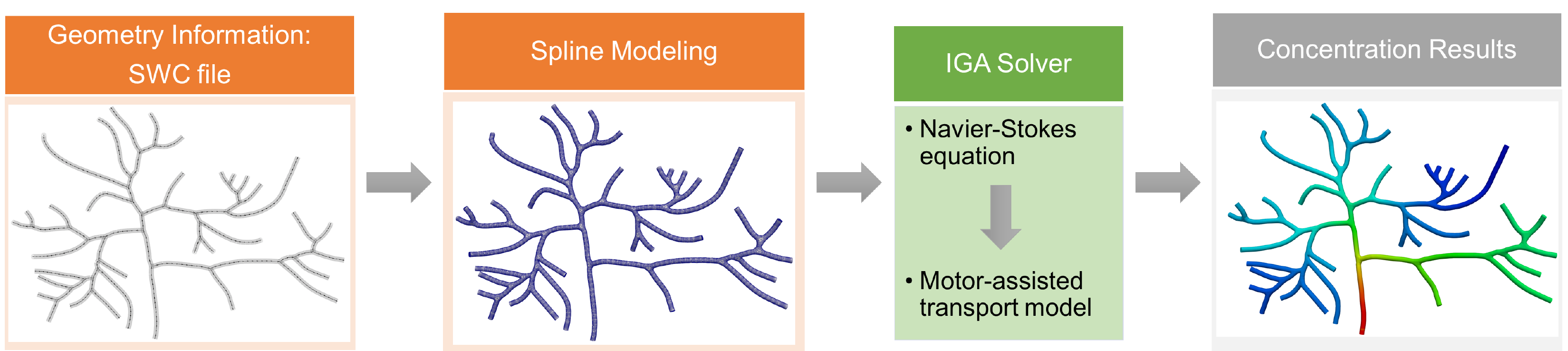}
    \caption{An overview of the IGA-based neuron material transport simulation pipeline. Given the geometry information (skeleton and diameter) of the neurite network stored in a SWC file, the spline modeling module first generates all-hexahedral control mesh and builds volumetric splines over the mesh. Then, the IGA solver takes the volumetric splines as input and computes the dynamic concentration results.}
    \label{figS:IGAPipeline}
  \end{figure*}
  
  \begin{figure*}[!htb]
    \centering
    \includegraphics[width = \linewidth]{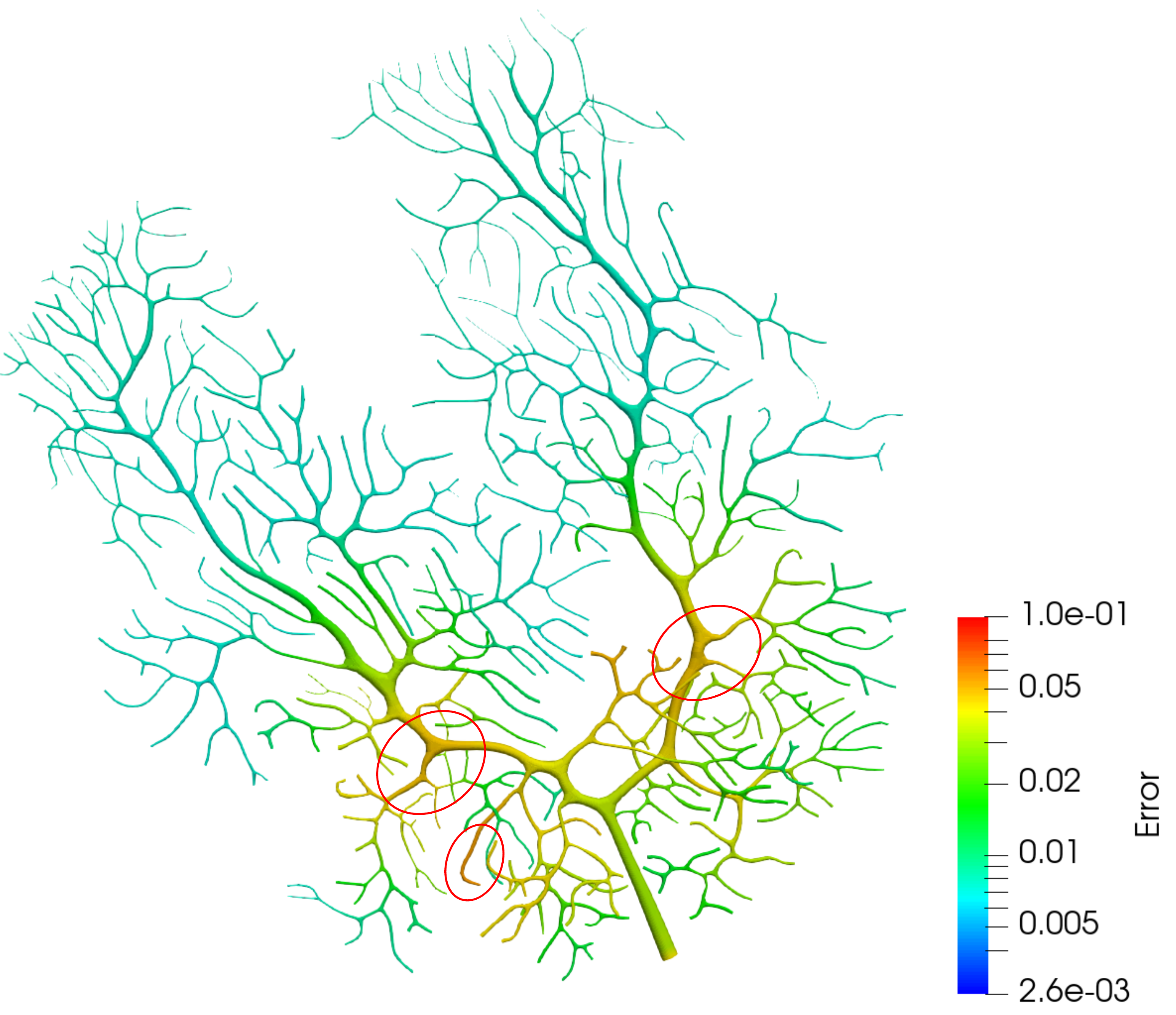}
    \caption{The prediction error of NMO\_00865 using the model trained with zebrafish neuron dataset. Logarithmic scale is used to highlight the distribution pattern. The regions with higher prediction error are labeled in red circles where the geometry shows high curvature or sharp radius change.}
    \label{figS:Result_Purkenje}
  \end{figure*}

\end{document}